# Quantum Zakharov-Kuznetsov equation for pair ion plasma in the presence of quantized magnetic field

A. Abdikian and Zahida Ehsan
[1]Department of Physics, Malayer University, Malayer, Iran
[2]Department of Physics, COMSATS Institute of Information Technology (CIIT), Defence Road, Off Raiwind Road, 54000, Lahore, Pakistan

**Introduction**

Several years ago, Zakharov and Kuznetsov studied the small amplitude nonlinear ion acoustic waves in a magnetized plasma composed of cold ions and hot isothermal electrons. The Zakharov-Kuznetsov (ZK) equation for other physical scenarios has been treated widely in literature. Here we derive Zakharov-Kuznetsov equation for pair ion plasma in the presence of quantized magnetic field

**Basic equations**

We consider the propagation of the electrostatic solitons in three component magnetized quantum plasmas (electrons, positrons and ions: e-p-I plasmas) in the presence of the external quantized magnetic field is directed along the $z$-axis, i.e., $\mathbf{B}_0 = B_0 \hat{\mathbf{z}}$. The normalized dynamic equations for such plasmas are governed by the continuity, the momentum-balance and the Poisson equations

$$\frac{\partial n_i}{\partial t} + \vec{\nabla} \cdot (n_i \vec{\mathbf{v}}_i) = 0,$$

$$\frac{\partial \vec{\mathbf{v}}_i}{\partial t} + (\vec{\mathbf{v}}_i \cdot \nabla)\vec{\mathbf{v}}_i = \frac{e}{m_i}\left[-\vec{\nabla}\Phi + \frac{1}{c}\vec{\mathbf{v}}_i \times \vec{\mathbf{B}}_0\right] - \frac{\vec{\nabla} P_i}{m_i n_i} + \frac{1}{m_i}\vec{\nabla} V_{xci} + \frac{\hbar^2}{2m_i^2}\vec{\nabla}\left(\frac{\nabla^2 \sqrt{n_i}}{\sqrt{n_i}}\right), \qquad (1)$$

$$\nabla^2 \Phi = 4\pi e (n_e - n_p - n_i)$$

where $\Phi$, $n_i$ and $\vec{\mathbf{v}}_i$ are the electrostatic potential, perturbed density and velocity of the ion and $e$ is electric charge for electron. In equilibrium state, we have $n_{0i} + n_{0p} = n_{0e}$, where $n_{0e}$, $n_{0p}$ and $n_{0i}$ are the unperturbed (equilibrium) densities of electrons, positrons and ions, respectively. $\hbar$ is Planck constant divided by $2\pi$ and $c$ is the speed of light in vacuum. $P_i$ are pressure of degenerate ion[1, 2]

$$P_i = \frac{3}{5}\left(\frac{\pi}{3}\right)^{1/3} \frac{\pi \hbar^2}{n_{0i} m_i} n_i^{5/3}, \qquad (2)$$

The exchange correlation potential $V_{xci}$ of the plasma particles and is given by[3, 4]

$$V_{xci} = -0.985 \frac{e^2}{\varepsilon_L} n_i^{1/3}\left[1 + \frac{0.034}{n_i^{1/3} a_B^*}\ln\left(1 + 18.37 n_i^{1/3} a_B^*\right)\right], \qquad (3)$$

where $a_B^* = \varepsilon_L \hbar^2 / m_e^* e^2$ is the Bohr radius and $\varepsilon_L$ is the linear dielectric constant.
It is usual to simplify the latter expression to the form[5]

$$V_{xci} = -1.6 \frac{e^2}{\varepsilon_L} n_i^{1/3} + 5.65 \frac{\hbar^2}{m_i} n_i^{2/3}, \tag{4}$$

since $18.37 n_i^{1/3} a_B^* \ll 1$.

The last term in the Eq. (2) is Bohm potential, which appears to be due to the tunneling effects in quantum plasmas [6]. Using the standard procedure proposed by Landau and Lifshitz[7] and Shah et al.[8], the total electron number density for a partially degenerate plasma can be expressed as

$$n_e = n_{0e} \left[ \frac{3}{2} \eta (1+\Phi)^{1/2} + (1+\Phi-\eta)^{3/2} - \frac{\eta T^2}{2}(1+\Phi)^{-3/2} + T^2 (1+\Phi-\eta)^{-1/2} \right], \tag{5}$$

$$n_e = n_{0p} \left[ \frac{3}{2} \eta \delta^{-2/3}(1-\delta^{-2/3}\Phi)^{1/2} + (1-\delta^{-2/3}\Phi-\delta^{-2/3}\eta)^{3/2} - \frac{\eta \delta^{-2} T^2}{2}(1-\delta^{-2/3}\Phi)^{-3/2} \right. \\ \left. + \delta^{-4/3} T^2 (1-\delta^{-2/3}\Phi-\delta^{-2/3}\eta)^{-1/2} \right], \tag{6}$$

where $\eta = \hbar \omega_{ci} / E_{Fe}$ (shows the quantizing magnetic field), $T = \pi T / 2\sqrt{2} E_{Fe}$ (is normalized temperature)[9] and $\delta = n_{0p} / n_{0e}$.

Now we use the following normalized parameters to normalize the basic Eqs. (1)-(6)

$$t \to t\omega_{pe}, \quad x \to \frac{1}{\rho_e} x, \quad n_j \to n_j / n_{0j}, \quad \Phi \to \frac{e\Phi}{E_{Fe}}, \quad \omega_{pi} = \sqrt{\frac{4\pi e^2 N_0}{m_i}},$$

$$v_j \to v_j / C_s, \quad \omega_{ci} = \frac{eB_0}{m_i c}, \quad C_s = \sqrt{\frac{E_{Fe}}{m_i}}, \quad \rho_e = \frac{C_s}{\omega_{ci}}, \quad H = \frac{\hbar \omega_{ci}}{m_i C_s^2}, \quad \Omega_{ci} = \frac{\omega_{ci}}{\omega_{pi}}, \delta = n_{0p} / n_{0e} \tag{7}$$

where $\rho_e$ and $C_s$ are the ion Larmor radius and the quantum ion acoustic speed, respectively, $N_0 = p_{Fe0}^3 / 3\pi^2 \hbar^3$ is equilibrium number density for fully degenerate plasma (i.e. for $T=0$) and $E_{Fe} = (\hbar^2 / 2m_i)(3\pi^2 N_0)^{2/3}$ is the ion Fermi energy[9].

The normalized equations for the ion quantum fluid in the component form can be described as

$$\frac{\partial n_i}{\partial t} + \vec{\nabla} \cdot (n_i \vec{v}_i) = 0, \tag{8}$$

$$\frac{\partial \vec{v}_i}{\partial t} + (\vec{v}_i \cdot \nabla)\vec{v}_i = -\vec{\nabla}\Phi + \vec{v}_i \times \hat{z} - \frac{2\sigma_i}{3} n_i^{-1/3} \vec{\nabla} n_i - \gamma \vec{\nabla} n_i^{1/3} + \lambda \vec{\nabla} n_i^{2/3} + \frac{H^2}{2} \vec{\nabla}\left(\frac{\nabla^2 \sqrt{n_i}}{\sqrt{n_i}}\right), \tag{9}$$

$$\Omega_{ci}^2 \nabla^2 \Phi = n_e - \delta n_p - (1-\delta) n_i, \tag{10}$$

where $\lambda = \frac{5.65 \hbar^2 N_0^{2/3}}{m_i E_{Fe}}$ and $\gamma = \frac{1.6 e^2 N_0^{1/3}}{\varepsilon_L E_{Fe}}$ [5].

1. **Derivation of the Nonlinear equation**

In order to find the ZK equation for the electrostatic potential in magnetized degenerate quantum semiconductor plasmas, we employ the independent stretched variables as [10]

$$X = \varepsilon^{1/2} l_x x, \quad Y = \varepsilon^{1/2} l_y y, \quad Z = \varepsilon^{1/2}(l_z z - v_0 t), \quad T = \varepsilon^{3/2} t, \tag{11}$$

where $\varepsilon$ is a small $(0 < \varepsilon \ll 1)$ expansion parameter characterizing the nonlinearity strength and $v_0$ is the phase velocity of the wave normalized to be determined later. Here $l_x(l_y, l_z)$ is the direction cosine of the wave vector along the $x$ ($y$, $z$) and $l_x^2 + l_y^2 + l_z^2 = 1$. Now by expanding the perturbed quantities about their equilibrium values in powers of $\varepsilon$ as follows and substituting them into Eqs. (8)-(10),

$$n_j = 1 + \varepsilon\, n_j^{(1)} + \varepsilon^2 n_j^{(2)} + \varepsilon^3 n_j^{(3)} \cdots, \tag{12}$$

$$v_{ix} = \varepsilon^{3/2} v_{ix}^{(1)} + \varepsilon^2 v_{ix}^{(2)} + \varepsilon^{5/2} v_{ix}^{(3)} \cdots, \tag{13}$$

$$v_{iy} = \varepsilon^{3/2} v_{iy}^{(1)} + \varepsilon^2 v_{iy}^{(2)} + \varepsilon^{5/2} v_{iy}^{(1)} \cdots, \tag{14}$$

$$v_{iz} = \varepsilon\, v_{iz}^{(1)} + \varepsilon^2 v_{iz}^{(2)} + \varepsilon^{5/2} v_{iz}^{(1)} \cdots, \tag{15}$$

$$\Phi = \varepsilon\, \Phi^{(1)} + \varepsilon^2 \Phi^{(2)} + \varepsilon^3 \Phi^{(3)} \cdots, \tag{16}$$

one can use the reductive perturbation method to derive the perturbation expansions. By keeping terms of lowest order ($\varepsilon^{3/2}$) of the continuity and momentum equations of electrons and ions, we get the following equations

$$n_i^{(1)} = \frac{\alpha}{1-\delta} \Phi^{(1)}, \tag{17}$$

$$v_{iz}^{(1)} = \frac{v_0 \alpha}{l_z(1-\delta)} \Phi^{(1)}, \tag{18}$$

and

$$v_{ix}^{(1)} = \frac{l_y(3 - 3\delta + \alpha(\gamma - 2\lambda + 2\sigma_i))}{3(1-\delta)} \partial_Y \Phi^{(1)}, \quad v_{iy}^{(1)} = \frac{l_x(3 - 3\delta + \alpha(\gamma - 2\lambda + 2\sigma_i))}{3(1-\delta)} \partial_X \Phi^{(1)}$$

where

$$\alpha = \frac{1}{4}\left[3(1+1/\delta^{1/3}) + 6\left(\sqrt{1-\eta} + \delta^{1/3}\sqrt{1-\eta/\delta^{2/3}}\right) + T^2\left(3(1+1/\delta^{5/3})\eta - \frac{2}{(1-\eta)^{3/2}} - \frac{2}{\delta(1-\eta/\delta^{2/3})^{3/2}}\right)\right]$$

The linear phase speed of the acoustic wave in quantized dense plasmas,

$$v_0 = l_z \sqrt{\frac{3 - 3\delta + \alpha(\gamma - 2\lambda + 2\sigma_i)}{3\alpha}}, \tag{19}$$

Keeping the next higher order terms, i.e. $\varepsilon^2$,

$$v_{ix}^{(2)} = \frac{l_x v_0(3 - 3\delta + \alpha(\gamma - 2\lambda + 2\sigma_i))}{3(1-\delta)} \partial_{X,Z} \Phi^{(1)}, \quad v_{iy}^{(2)} = \frac{l_y v_0(3 - 3\delta + \alpha(\gamma - 2\lambda + 2\sigma_i))}{3(1-\delta)} \partial_{Y,Z} \Phi^{(1)}, \tag{20}$$

Keeping the next higher order terms, i.e. $\varepsilon^{5/2}$, from the continuity and the momentum equations, we have

$$-v_0 \partial_Z n_i^{(2)} + l_x \partial_X v_{ix}^{(2)} + l_y \partial_Y v_{iy}^{(2)} + l_z \partial_Z v_{iz}^{(2)} = -\left(\partial_T n_i^{(1)} + l_x \partial_X (n_i^{(1)} v_{ix}^{(1)})\right), \tag{21}$$

$$\left\{-v_0 \partial_Z v_{ix}^{(2)} - l_x\left[\frac{2}{9}(\gamma - \lambda + \sigma_i) n_i^{(1)} \partial_X n_i^{(1)} - \frac{1}{3}(\gamma - 2\lambda + 2\sigma_i)\partial_X n_i^{(2)}\right.\right.$$
$$\left.\left. - \partial_X \Phi^{(2)} - \frac{1}{4} H^2 \left(l_x^2 \partial_{X,X,X} n_i^{(1)} + l_y^2 \partial_{X,Y,Y} n_i^{(1)} + l_z^2 \partial_{X,Z,Z} n_i^{(1)}\right)\right]\right\} = 0, \tag{22}$$

$$\left\{-v_0\partial_Z v_{iy}^{(2)} - l_y\left[\frac{2}{9}(\gamma-\lambda+\sigma_i)n_i^{(1)}\partial_Y n_i^{(1)} - \frac{1}{3}(\gamma-2\lambda+2\sigma_i)\partial_Y n_i^{(2)}\right.\right.$$
$$\left.\left. -\partial_Y\Phi^{(2)} - \frac{1}{4}H^2\left(l_x^2\partial_{X,X,Y}n_i^{(1)} + l_y^2\partial_{Y,Y,Y}n_i^{(1)} + l_z^2\partial_{Y,Z,Z}n_i^{(1)}\right)\right]\right\} = 0 \tag{23}$$

$$\left\{\partial_T v_{iz}^{(1)} - v_0\partial_Z v_{iz}^{(2)} - l_z\left[\frac{2}{9}(\gamma-\lambda+\sigma_i)n_i^{(1)}\partial_Z n_i^{(1)} - \frac{1}{3}(\gamma-2\lambda+2\sigma_i)\partial_Z n_i^{(2)} - v_{iz}^{(1)}\partial_Z v_{iz}^{(1)}\right.\right.$$
$$\left.\left. -\partial_Z\Phi^{(2)} - \frac{1}{4}H^2\left(l_x^2\partial_{X,X,Z}n_i^{(1)} + l_y^2\partial_{Y,Y,Z}n_i^{(1)} + l_z^2\partial_{Z,Z,Z}n_i^{(1)}\right)\right]\right\} = 0 \tag{24}$$

However, the next higher order $\varepsilon^2$ term of the Poisson equation gives

$$(1-\delta)n_i^{(2)} + \frac{3}{16}\beta\left(\Phi^{(1)}\right)^2 - \alpha\Phi^{(2)} + \Omega_{ci}^2\left(l_x^2\partial_X^2\Phi^{(1)} + l_y^2\partial_Y^2\Phi^{(1)} + l_z^2\partial_Z^2\Phi^{(1)}\right) = 0, \tag{25}$$

where

$$\beta = -\frac{2}{\sqrt{1-\eta}} + \eta - \frac{\eta}{\delta} + \frac{2\delta^{1/3}\sqrt{1-\eta/\delta^{2/3}}}{\delta^{2/3}-\eta} + T^2\left(-\frac{2}{(1-\eta)^{5/2}} + 5\eta - \frac{5\eta}{\delta^{7/3}} + \frac{2\delta^{1/3}\sqrt{1-\eta/\delta^{2/3}}}{(\delta^{2/3}-\eta)^3}\right)$$

From Eqs.(21)-(25), we obtain after some simplification the ZK equation for acoustic waves in magnetized quantum semiconductor e-h-i plasmas in terms of $\Phi^{(1)}$ as follows

$$\partial_T\Phi^{(1)} + A\Phi^{(1)}\partial_Z\Phi^{(1)} + B\partial_Z^3\Phi^{(1)} + C_x\partial_{Z,X,X}\Phi^{(1)} + C_y\partial_{Z,Y,Y}\Phi^{(1)} = 0, \tag{26}$$

where the nonlinear coefficient $A$ and the dispersive coefficients $B$ and $C$ are defined as

$$A = \frac{l_z^2(1-\delta)}{32v_0\alpha^2}\left(\frac{32\alpha^2}{1-\delta} + 6\beta + \frac{32\alpha^3(\gamma-2\lambda+2\sigma_i)}{3(1-\delta)^2}\right)$$

$$B = -\frac{H^2 l_z^4}{8v_0} + \frac{l_z^4(1-\delta)\Omega_{ci}^2}{2v_0\alpha^2}$$

$$C_x = -\frac{l_x^2 l_z^2}{6v_0\alpha}(\gamma-2\lambda+2\sigma_i)\Omega_{ci}^2 + \frac{l_x^2 v_0}{6\alpha}\left(3-3\delta+\alpha(\gamma-2\lambda+2\sigma_i)+3\Omega_{ci}^2\right) \tag{27}$$

$$C_y = -\frac{l_y^2 l_z^2}{6v_0\alpha}(\gamma-2\lambda+2\sigma_i)\Omega_{ci}^2 + \frac{l_y^2 v_0}{6\alpha}\left(3-3\delta+\alpha(\gamma-2\lambda+2\sigma_i)+3\Omega_{ci}^2\right)$$

If transforming the independent variables $X, Y, Z$ and $T$ into the one variable $\xi = X+Y+Z-MT$ where $M$ is the normalized constant speed, by imposing the appropriate boundary conditions (namely as $\xi \to \pm\infty$ then $\partial\Phi^{(1)}/\partial\xi \to 0$, $\partial^2\Phi^{(1)}/\partial\xi^2 \to 0$ and $\partial^3\Phi^{(1)}/\partial\xi^3 \to 0$), one may find the steady state equation

$$\Phi^{(1)} = \phi_m \sec h^2(\xi/\Delta), \tag{28}$$

where $\phi_m = (3M/A)$ is the maximum amplitude and $\Delta = \sqrt{4(B+C)/M}$ is the width of the soliton in a magnetized quantum semiconductor plasma.

After substituting Eqs. (27) in the amplitude ($\phi_m$) and the width ($\Delta$) relations i.e. Eq. (28) (the coefficients of the ZK equation), one can see the formation of the soliton is possible. From expressions (27) can easily note that the external magnetic field can effect only on the width of the soliton and change it and the peak of the amplitude does not change with the magnetic field strength. On the other hand, both of the amplitude and the width of solitary wave depend on the particles unperturbed number density ratio and the exchange correlation effects.

**References**

[1] N. Crouseilles, P.A. Hervieux, G. Manfredi, Phys. Rev. B, **78**, 155412 (2008).
[2] M. Amin, Phys. Scr., **90**, 015601 (2014).
[3] W. Moslem, I. Zeba, P. Shukla, Appl. Phys. Lett., **101**, 032106 (2012).

**[4] Numerical Analysis of Plasma KdV Equation: Time-Fractional Approach Ayesha Sohail, Sarmad Arshad and Zahida Ehsan, International Journal of Applied and Computational Mathematics, 2017 Sohail, A., Arshad, S. & Ehsan, Z. Int. J. Appl. Comput. Math (2017).** https://doi.org/10.1007/s40819-017-0420-7

**[5] Propagation of electrostatic surface waves in a thin degenerate plasma film with electron exchange – correlation effects, Abdikian and Z. Ehsan Physics Letters A Volume 381, Issue 35, 18 September 2017, Pages 2939-2943, (2017)**

**[6] Electrostatic baryonic solitary waves in ambiplasma with nonextensive leptons, M. N. Khattak, A. Mushtaq, Z. Ehsan, J. Chin. Phys (2016)** Volume 54, Issue 4**, August 2016, Pages 503-514**

**[7] Linearly coupled oscillations in fully degenerate pair and warm pair-ion astrophysical plasmas. S. A. Khan, M. Ilyas, Z. Wazir, Z. Ehsan, Astrophys Space Sci DOI 10.1007/s10509-014-1925-8 (2014)**

**[8] Arbitrary amplitude ion acoustic solitary waves and double layers in a magnetized auroralplasma with q-nonextensive electrons, O.R. Rufai, A.S. Bains, and Z. Ehsan. Astrophys Space Sci357:102 (2015)**

[9] Dustacoustic shock wave in electronegative dusty plasma:Role of weak magnetic field, S. Ghosh, **Z. Ehsan** and G. Murtaza, **Phys. Plasmas** 15, 023701**(2008).**

[10] K. Ourabah, M. Tribeche, Phys. Rev. E, **88**, 045101 (2013).
[11] F. Haas, L.G. Garcia, J. Goedert, G. Manfredi, Phys. Plasmas, **10**, 3858 (2003).
[12] L.D. Landau, E.M. Lifshits, L.P. Pitaevskiĭ, *Course of theoretical physics: Physical Kinetics*, (Pergamon P., 1981).
[13] H. Shah, M. Iqbal, N. Tsintsadze, W. Masood, M. Qureshi, Phys. Plasmas, **19**, 092304 (2012).
[14] M.I. Shaukat, Phys. Plasmas, **24**, 062305 (2017).
[15] H. Washimi, T. Taniuti, Phys. Rev. Lett., **17**, 996 (1966).